\documentclass[prl,twocolumn,superscriptaddress,preprintnumbers,aps,showpacs,amsmath,amssymb]{revtex4}

\usepackage{graphicx}
\usepackage{dcolumn}
\usepackage{bm}

\begin{document}

\title{Spontaneous motion of a droplet coupled with a chemical wave}

\author{Hiroyuki Kitahata\footnote{Corresponding author. E-mail: kitahata@physics.s.chiba-u.ac.jp.}}
\affiliation{Department of Physics, Graduate School of Science, Chiba University, Chiba 263-8522, Japan}
\affiliation{PRESTO, JST, Saitama 332-0012, Japan}

\author{Natsuhiko Yoshinaga}
\affiliation{Fukui Institute for Fundamental Chemistry, Kyoto University, Kyoto 606-8103, Japan}

\author{Ken H. Nagai\footnote{Present address: Department of Physics, Graduate School of Science, The University of Tokyo, Tokyo 133-0033, Japan.}}
\affiliation{Division of Advanced Sciences, Ochadai Academic Production, Ochanomizu University, Tokyo 112-8610, Japan}

\author{Yutaka Sumino\footnote{Present address: Faculty of Education, Aichi University of Education, Aichi 448-8542, Japan.}}
\affiliation{Department of Applied Physics, The University of Tokyo, Tokyo 113-8656, Japan}

\begin{abstract}
We propose a novel framework for the spontaneous motion of a droplet
coupled with internal dynamic patterns generated in a
reaction-diffusion system. The spatio-temporal order of the chemical
reaction gives rise to inhomogeneous surface tension and results in
self-propulsion driven by the surrounding flow due to the Marangoni
effect. Numerical calculations of internal patterns together with
theoretical results of the flow fields at low Reynolds number well
reproduces the experimental results obtained using a droplet of
Belousov-Zhabotinsky (BZ) reaction medium.
\end{abstract}


\maketitle

Spatio-temporal patterns are widely seen in living systems; target, spiral, stripe, and dot patterns have been observed at various scales from the interior of a cell to a swarm of cells. Most of studies have been focused on patterns at larger scales, which can be successfully reproduced using reaction-diffusion dynamics~\cite{pattern}. In contrast, it is only recently that internal patterns in a single cell have been visualized. These patterns are expected to relate to cellular functions; examples include calcium ions for signal transduction~\cite{Mathphys}, Min proteins for cell division~\cite{Min}, and actin cytoskelton for mechanical properties~\cite{Vicker}.  Although pattern formation in a cell is expected to be analyzed in the framework of a reaction-diffusion system, as demonstrated in {\it in vitro} experiments \cite{Min}, sufficient understanding on the connection between pattern formation and cellular function is awaited.  In this paper, we focus motility, as a typical aspect of cellular functions, arising from internal patterns.

Several artificial systems imitating cell motility have been proposed
as self-propelled particles~\cite{self-propulsion}.
Although no external force is exerted on the particles (force-free
condition), the motion is induced by the asymmetric distribution of
an electric field, concentration of chemicals, temperature, and so on.
These asymmetric distributions are either {\it a priori} embedded in
the asymmetry of the surface properties of the self-propelled
objects~\cite{asymmetry}, or {\it a
posteriori} created by nonlinear effects; motion itself destabilizes
symmetric distribution, for instance, through advective
flow~\cite{bifurcation}.
In both cases, however, most studies have focused on motion under
steady distributions.

In order to understand the dynamic features of cell motility, a system
connecting dynamic pattern with motion is desirable. In fact, experimental and numerical evidence of chemo-mechanical coupling in such systems has been demonstrated~\cite{Kitahata-JCP,Yeomans}. 
In this letter, we propose a theoretical framework for a chemical system exhibiting self-organized
patterns, leading to spontaneous motion.  We consider that the Marangoni effect is suitable for this purpose as it has been shown to
drive an object under force-free conditions by an inhomogeneous
interfacial tension arising from a gradient in chemical concentration~\cite{Marangoni,interface}. In our system, the
energy supply and consumption can generate a pattern in a droplet
through nonlinear chemical kinetics, and the pattern at the interface
of the droplet creates inhomogeneous interfacial tension. This generates
a flow surrounding the droplet, resulting in motion.

We discuss a spherical droplet of incompressible fluid with a radius of $R$ in another fluid. We consider the Stokes equation under the approximation of low Reynolds number~\cite{Brenner}. The inertia term $\rho {\rm d}{\bf v}/{\rm d}t$ is also neglected where $\rho$ is fluid density. We will justify these assumptions later using experimental values. The concentration of chemical species, which determines the interfacial tension, obeys the reaction-diffusion-advection equation. The sets of governing equations are described as 
\begin{eqnarray}
&\rho \partial_t {\bf v} = -\nabla p + \eta \nabla^2 {\bf v} = 0, \label{Stokes}&\\
&\nabla \cdot {\bf v} = 0, \label{incompressible}&\\
&\partial_t {\bf c} + \left({\bf v} \cdot \nabla \right) {\bf c} = {\bf F}\left({\bf c}\right) + \nabla \left(D \nabla {\bf c}\right),&\label{RDA}
\end{eqnarray}
where ${\bf v}$, $p$, and $\eta$ are the flow velocity, pressure, and viscosity of the fluid, respectively. ${\bf c}({\bf r})$ is a vector composed of concentrations of chemicals, ${\bf F}({\bf c})$ corresponds to the reaction kinetics, and $D$ is a diffusion coefficient. 

The flow velocity is coupled with the chemical reaction through interfacial tension at the droplet interface. This coupling is described as a force balance in a tangential direction at an interface~\cite{interface,Young}:
$\left. \sigma_{r\theta}^{\rm (i)}\right|_{r=R} = \left. \sigma_{r\theta}^{\rm (o)}\right|_{r=R} + (\partial \gamma/\partial \theta) / R,$
where $\sigma$ is the stress tensor, and $\gamma(\theta)$ is the interfacial tension profile, which is dependent on ${\bf c}$.

\begin{figure}
\includegraphics{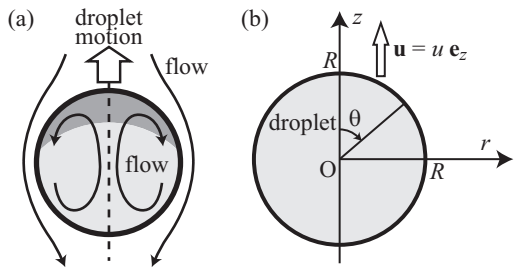}
\caption{(a) Schematic representation of the system under consideration. The broken line corresponds to the symmetry axis. (b) Setting of the coordinates in the droplet system.}
\label{fig1}
\end{figure}

For simplicity, we only consider the axisymmetric system schematically shown in Fig. \ref{fig1}(a). Since the boundary condition is given at the droplet interface, it is convenient to set the spherical coordinates so that the droplet is fixed, as shown in Fig. \ref{fig1}(b). We have to adopt the boundary condition that $v_r^{({\rm i})} = v_r^{({\rm o})} = 0$ and $v_\theta^{({\rm i})} = v_\theta^{({\rm o})}$ at $r = R$ and ${\bf v} \rightarrow {\bf v}_0 = -{\bf u}$ as $r \rightarrow \infty$, where $r$ is the distance from the droplet center, and ${\bf u}$ is the droplet velocity in a laboratory system. The superscripts ``(i)'' and ``(o)'' correspond to the fluids inside and outside of the droplet, respectively. From the symmetric property, we aligned ${\bf u}$ with positive $z$; i.e., ${\bf u} = u {\bf e}_z$. 
 We also have to consider the condition that the force exerted on the droplet, ${\bf f}$, is zero; i.e., ${\bf f} = \int {\rm d}S \, {\bf n} \cdot \left. \sigma^{\rm (o)} \right|_{r=R} = {\bf 0}.$

By solving Eqs. (\ref{Stokes}) and (\ref{incompressible}) with the above conditions, the droplet velocity is calculated as
\begin{equation} 
u = -\frac{2}{9\eta^{\rm (i)} + 6\eta^{\rm (o)}} \Gamma_1 \equiv -\alpha \Gamma_1, \label{equ}
\end{equation}
where $\Gamma_1$ is the first-mode coefficient of the Legendre expansion of $\gamma(\theta)$~\cite{Young,JCIS}:
$\gamma(\theta) = \sum_{n=0}^\infty \Gamma_n P_n(\cos \theta)$,
where $P_n$ is the Legendre polynominal of order $n$. The higher modes ($n \geq 2$) contribute only to deformation and are independent to the translational motion for small deformation.
The detailed derivation on the solutions of the Stokes equation is shown in Supplemental Materials (SM)~\cite{SM}.

\begin{figure}
\includegraphics{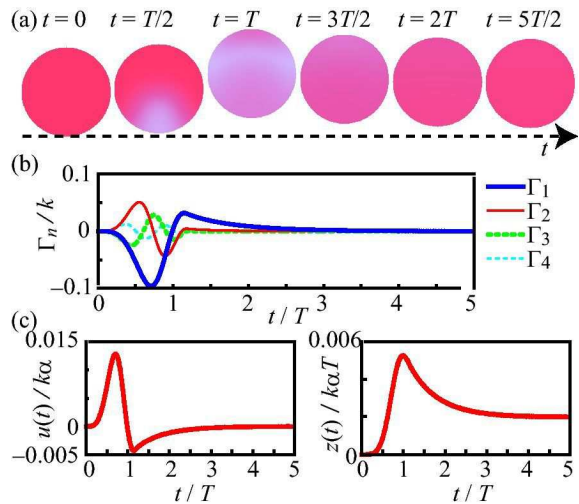}
\caption{(Color online) Numerical calculations for a single pulse. (a) Snapshots of $V$ at the cross section and displacement of a droplet. Blue (light gray) and red (dark gray) colors correspond to higher and lower $V$, respectively. The displacement is shown as the vertical position. At $t = 0$, a chemical wave was initiated by setting $U=1$ at a certain region in the lower half of the droplet. (b) $\Gamma_n \, (1 \leq n \leq 4)$ against time. (c) Velocity $u$ and the position $z$ of the droplet against time. The parameters: $R = 125$, $D = 1600$.}
\label{resnum}
\end{figure}

We focus on a specific system using the Belousov-Zhabotinsky (BZ) reaction, which is common in experimental systems that can exhibit spatio-temporal pattern formation, such as target and spiral patterns in a two-dimensional system~\cite{BZ} and a scroll ring in a three-dimensional system~\cite{scroll}. As a mathematical model for the BZ reaction, we adopt the Oregonator~\cite{Oregonator}, which is widely used due to its reliability as demonstrated in physico-chemical discussion on the elemental processes of chemical reactions. In the Oregonator, ${\bf c}$ is composed of the two variables $U$ and $V$, which corrrepond to the concentrations of HBrO$_2$ and oxidized catalyst, also referred to as the activator and inhibitor, respectively. ${\bf F}({\bf c})$ is described as:
\begin{equation}
{\bf F} \left({\bf c}\right) = {\bf F} \! \left( \!\! \begin{array}{c} U \\ V \end{array} \!\! \right) = \left( \!\! \begin{array}{c} \displaystyle{\frac{1}{\epsilon} \left[U(1-U) - fV\frac{U-q}{U+q}\right]} \\ U-V \end{array} \!\! \right),
\end{equation}
where $\epsilon$, $q$, and $f$ are the parameters that determine the characterisitcs of the BZ reaction. The Oregonator model is nondimensionalized, so that the time unit in the calculation is set as $T$.

The reaction-diffusion-advection equation (Eq.~(\ref{RDA})) is numerically solved inside a droplet, which  corresponds to the experimental conditions in which a droplet of BZ reaction medium (BZ droplet) is inside another fluid. Here, by taking the axisymmetry into consideration, the calculations can be performed in a two-dimensional field . We adopt a monotonous increasing function for the interfacial tension against $V$:
\begin{equation}
\gamma (V) = \gamma_0 + k V. \label{eq-gamma}
\end{equation}
This assumption is supported by the experimental measurements; the air/water interfacial tension in the ferroin-catalyzed BZ reaction medium is higher in the oxidized state than that in the reduced state~\cite{BZ-surface,inomoto}. In addition, Marangoni flow was observed at the oil/water interface, as well as the air/water interface~\cite{Kitahata-JCP}. Note that the coefficient may include the effect due to the heat generation by reaction, which only reduces its absolute value.

The parameters for the Oregonator model were set as $q= 0.001$, $\epsilon = 0.05$, and $f = 2.5$, which correspond to the excitable condition, in which a chemical wave propagates only from the initiated point. In the numerical calculation, the velocity field was calculated by summing the modes of $n=1$ to $8$. We confirmed that the cut-off modes did not critically affect the numerical results.   

\begin{figure}
\includegraphics{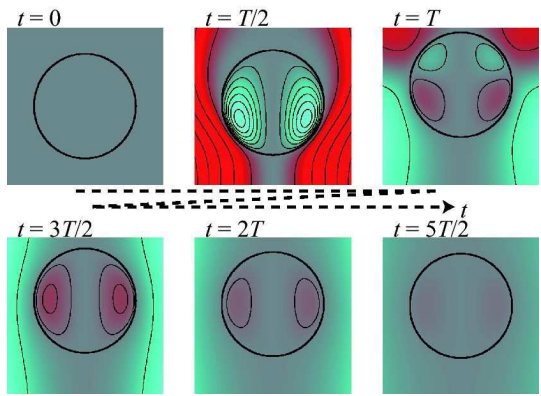}
\caption{(Color online) Snapshots of streamlines and stream function at the cross section corresponding to the snapshots in Fig. \ref{resnum} (a). Red (dark gray) and blue (light gray) colors correspond to positive and negative values of the stream function, respectively~\cite{SM}. The direction of the rolls in each droplet is shown by red (dark gray) or blue (light gray) coloring (clockwise: red (dark gray) in the left hemisphere and blue (light gray) in the right one, counterclockwise: blue (light gray) in the left hemisphere and red (dark gray) in the right one).}
\label{fig-flow}
\end{figure}

We performed numerical calculations under two typical sets of conditions; one is a single pulse that corresponds to a target pattern, and the other is a scroll ring~\cite{scroll}, which is a natural extension of a spiral wave in a three-dimensional medium.

Figure \ref{resnum} shows the results of numerical calculations for a single pulse. At $t=0$, a chemical wave was initiated in the lower half of the droplet, and then a single chemical wave propagated isotropically. Figure \ref{resnum} (a) shows snapshots of $V$ in a cross section of the droplet, and Fig. \ref{resnum} (b) and (c) are the time series of $\Gamma_n \, (1 \leq n \leq 4)$, velocity and position of the droplet, respectively. The chemical wave arose from the lower part of the droplet, and propagated in a circular shape. When the wave reached the lower boundary of the droplet, the droplet began to move upward. When the chemical wave reached the upper boundary, the droplet moved back in the opposite direction. Nevertheless, there was a net motion in a cycle (see Fig. \ref{resnum}(c) and movies in SM \cite{SM}.) Figure \ref{fig-flow} shows snapshots of streamlines in a cross section inside a droplet from the droplet-fixed frame. The direction of the convective rolls was inverted at the moment the droplet moved back.

\begin{figure}
\includegraphics{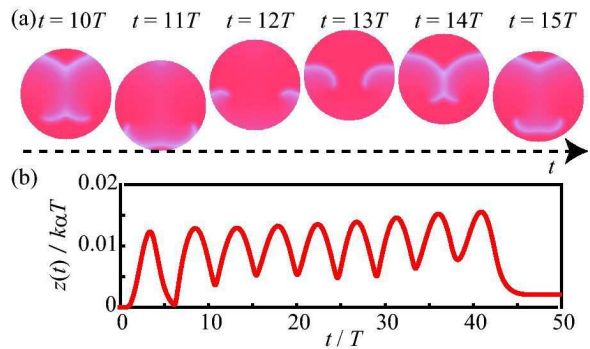}
\caption{(Color online) Numerical calculations for a scroll wave. (a) Snapshots of $V$ at the cross section and displacement of a droplet, shown in the same manner as in Fig. \ref{resnum}. At $t = 0$, a chemical wave was initiated by setting $U=1$ at a certain region in the lower half of the droplet, and the chemical wave at the central region of the droplet was omitted at $t = 2.5T$. (b) Time series of the position of the droplet. Parameters: $R = 125$, $D = 60$.}
\label{resnum-scroll}
\end{figure}

In the numerical calculation for a scroll ring, the pattern was prepared using the following standard procedure; a chemical wave was first initiated, and a part of the chemical wave was omitted by setting $U$ and $V$ as steady-state values at $t = 2.5T$.  The numerical results for the snapshots and time series of the position of the droplet are shown in Fig. \ref{resnum-scroll}. The droplet moved back and forth, gradually moving upward, coupling with the development of the scroll ring (see the movie in SM~\cite{SM}). In Fig. \ref{resnum-scroll}(a), the chemical wave first reached the lower half of the droplet, which lead to upward motion. As the point of contact moved toward the upper half of the droplet, downward motion resulted. The droplet finally stopped, since the filament (alignment of the phase singularity points) shrank and disappeared due to its intrinsic instability~\cite{scroll-instablity}.

\begin{figure}
\includegraphics{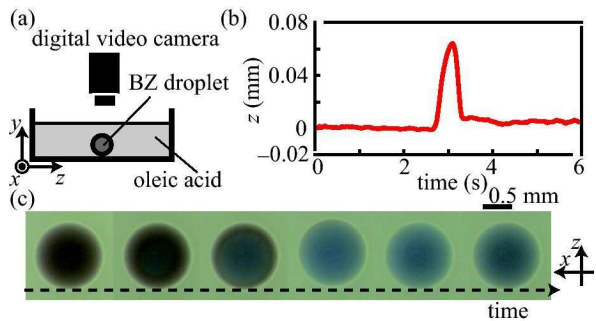}
\caption{(Color online) Experiments on the spontaneous motion of a BZ droplet. (a) Schematic representation of the experimental setup. (b) Time change in the center of mass of the BZ droplet. (c) Snapshots in a horizontal plane ($x$-$z$ plane) at every 1 s. Red (dark gray) and blue (light gray) regions correspond to the oxidized and reduced states in the BZ reaction, respectively.}
\label{fig_ex}
\end{figure}

In order to compare the theoretical and experimental results, we undertook experiments on spontaneous motion of a BZ droplet inside an oil phase as schematically shown in Fig. \ref{fig_ex}(a). We put a droplet of BZ medium with a volume of 1 $\mu$l into an oil phase (oleic acid), in a petri dish made of polytetrafluoroethylene to prevent the droplet from coming into contact with the bottom of the petri dish. We then observed the BZ droplet from above using a digital video camera (DCR-HC62, Sony, Japan) equipped with a close-up lens (CM-3500, Raynox, Japan). The composition of the BZ reaction medium was $[{\rm NaBrO}_3] = 0.3 \,{\rm M}$, $[{\rm H}_2{\rm SO}_4] = \,0.6 {\rm M}$, $[{\rm CH}_2({\rm COOH})_2] = \,0.1 {\rm M}$, $[{\rm NaBr}] = 30 \,{\rm mM}$ and $[{\rm Fe}({\rm phen})_3{\rm SO}_4] = 5 \,{\rm mM}$.

The experimental results are summarized in Fig. \ref{fig_ex}(b) and (c). The BZ droplet moves in a horizontal plane. The axis of the position in (b) is set so that the upper direction in (c) is positive. As the BZ reaction medium was prepared in an oscillatory condition, a chemical wave was initiated spontaneously from the point determined stochastically.  In Fig. \ref{fig_ex}(c), a chemical wave was initiated in the lower half of the droplet. When the propagating chemical wave touched the lower interface, the droplet began to move upwards. It then moved back in the opposite direction when the chemical wave propagated over the entire droplet. 

From the experimental results, the maximum droplet velocity was measured as $u_{\rm max} = 0.1 \, {\rm mm} \, {\rm s}^{-1}$. The viscosity of the BZ medium, $\eta^{\rm (i)}$, is $\sim \, 10^{-3}\, {\rm kg} \,{\rm m}^{-1}\,{\rm s}^{-1}$ and that of oleic acid, $\eta^{\rm (o)}$, is $\sim \, 10^{-2}\, {\rm kg} \,{\rm m}^{-1}\,{\rm s}^{-1}$. The Reynolds number in the present experimental system is calculated with the kinetic viscosity, $\nu$, as $Re = u R / \nu \sim 0.1$, which shows that the Stokes approximation is applicable. The relaxation of the velocity field is characterized by $\tau_u = R^2 / \nu \sim 1 \, {\rm s}$, while the time scale of the wave propagation is $\tau_{\rm p} = R/ \sqrt{D/\epsilon} \sim 1 \, {\rm s}.$ With these values, we confirm that the time-derivative in Eq.~(\ref{Stokes}) does not qualitatively affect our results, but modifies Eq.~(\ref{equ}) within 15\% following the method in \cite{Laplace}. Therefore, to the sake of simplicity, we assume the flow both inside and outside of the droplet rapidly changes following the changes in chemical concentration. Using these values and the results of the numerical calculations, the maximum difference in the interfacial tension between the oxidized and reduced conditions is estimated as $\Delta \gamma \, \sim \, 0.01 \, {\rm mN} \, {\rm m}^{-1}$. Although it is difficult to directly measure the interfacial tension at the oil/water interface in this particular system, $\Delta \gamma$ at an air/water interface of the BZ medium is available~\cite{inomoto}. The value in the reference was ten times higher than our estimation for $\Delta \gamma$ at an oil/water interface. In general, the interfacial tension of an oil/water interface is lower than that of an air/water interface~\cite{Israelachvili}, and thus the present estimation seems reasonable.

In summary, we studied droplet motion coupled with internal dynamic pattern formation through an interfacial tension gradient. The Stokes equation was adopted for a spherical droplet, and the motion of the droplet was calculated, which well reproduced the experimental results. Although we have not so far experimentally succeeded in generating scrolling rings in a BZ droplet, we expect this motion will be realized by using photosensitive BZ reaction. A similar model could also be applied to other types of motion; for example, rotational and oscillatory motion. It could further be extended to the motion of a nonspherical droplet. 

The authors thank T.~Ohta, K.~Yoshikawa, M.~Sano, H.~Kori and T.~Ichino for their helpful discussion. Y.~S. and K.~H.~N. are supported by a JSPS fellowship for young scientists (No.21-3566 and No.23-1819 ). 


\begin{thebibliography}{99}

\bibitem{pattern}K.~Matsumoto, S.~Takagi, and T.~Nakagaki, Biophys. J. {\bf 94}, 2492 (2008); S.~Kondo and R.~Asai, Nature {\bf 380}, 678 (1996); J.~D.~Murray, {\it Mathematical Biology} (Springer-Verlag, Berlin, 1989); H.~Meinhardt and P.~A.~J.~de~Boer, Proc. Nat. Am. Soc. {\bf 98}, 14202 (2001); J.~J.~Tyson {\it et al.}, Physica D {\bf 34}, 193 (1989).

\bibitem{Mathphys}J.~Keener and J.~Sneyd, {\it Mathematical Physiology} (Springer-Verlag, New York, 1998).

\bibitem{Min}M.~Loose {\it et al.}, Science {\bf 320}, 789 (2008).

\bibitem{Vicker}M.~G.~Vicker, Biophys. Chem. {\bf 84}, 87 (2000).



\bibitem{self-propulsion}S.~J.~Ebbens and J.~R.~Howse, Soft Matter {\bf 6}, 726 (2010); W.~F.~Paxton {\it et al.}, Angew. Chem. Int. Ed. {\bf 45}, 5420 (2006).

\bibitem{asymmetry}W.~F.~Paxton {\it et al.}, J. Am. Chem. Soc. {\bf 126}, 13424 (2004); 
J.~R.~Howse {\it et al.} Phys. Rev. Lett. {\bf 99}, 048102 (2007);
H.~R.~Jiang, N.~Yoshinaga, and M.~Sano, {\it ibid} {\bf 105}, 268302 (2010).

\bibitem{bifurcation}F.~D.~dos~Santos and T.~Ondarcuhu, Phys. Rev. Lett. {\bf 75}, 2972 (1995);
Y.~Sumino {\it et al.}, {\it ibid} {\bf 94}, 068301 (2005);
U.~Thiele and E.~Knobloch, {\it ibid} {\bf 97}, 204501 (2006); 
K.~Nagai {\it et al.}, Phys. Rev. E {\bf 71}, 065301 (2005);
M.~Nagayama {\it et al.}, Physica D {\bf 194}, 151 (2004).

\bibitem{Kitahata-JCP}H.~Kitahata {\it et al.}, J. Chem. Phys. {\bf 116} 5666 (2002).

\bibitem{Yeomans}K.~Furtado, C.~M.~Pooley, and J.~M.~Yeomans, Phys. Rev. E {\bf 78}, 046308 (2008).

\bibitem{Marangoni}L.~E.~Scriven and C.~V.~Sternling, Nature {\bf 187}, 186 (1960).

\bibitem{interface}A.~A.~Nepomnyashchy, M.~G.~Velarde, P.~Colinet, {\it Interfacial Phenomena and Convection} (Chapman \& Hall/CRC, Boca Raton, 2002).

\bibitem{Brenner}J.~Happel and H.~Brenner, {\it Low Reynolds Number Hydrodynamics: with special applications to particulate media} (Prentice-Hall, Englewood Cliffs, 1965).


\bibitem{Young}N.~O.~Young, J.~S.~Goldstein, and M.~J.~Block, J. Fluid Mech. {\bf 6}, 350 (1959).

\bibitem{JCIS}M.~D.~Levan, J. Colloid. Interface Sci. {\bf 83}, 11 (1981).

\bibitem{SM}
See supplementary material at
http://cu.phys.s.chiba-u.ac.jp/~kitahata/BZdrop/.



\bibitem{BZ}
R.~Kapral and K.~Showalter, {\it Chemical Waves and Patterns} (Kluwer Academic, Dordrecht, 1995).

\bibitem{scroll}A.~T.~Winfree, Scinece {\bf 181}, 937 (1973).

\bibitem{Oregonator}J.~P.~Keener and J.~J.~Tyson, Physica D {\bf 21}, 307 (1986). 

\bibitem{BZ-surface}K.~Yoshikawa {\it et al.}, Chem. Phys. Lett. {\bf 211}, 211 (1993);
H.~Miike, S.~C.~M\"{u}ller, and B.~Hess, Phys. Rev. Lett. {\bf 61}, 2109 (1988);
M.~Diewald {\it et al.}, {\it ibid} {\bf 77}, 4466 (1996);
K.~Matthiessen, H.~Wilke, and S.~C.~M\"{u}ller, Phys. Rev. E {\bf 53}, 6056 (1996).

\bibitem{inomoto}O.~Inomoto {\it et al.} Phys. Rev. E {\bf 61}, 5326 (2000).


\bibitem{scroll-instablity}A.~T.~Winfree and W.~Jahnke, J. Phys. Chem. {\bf 93}, 2823 (1989).

\bibitem{Laplace}V.~Galindo {\it et al.} Micrograv. Sci. Technol. {\bf 7}, 234 (1994).

\bibitem{Israelachvili}J.~N.~Israelachvili, ``Intermolecular and surface forces'', (Academic Press, London, 1985).

\end{thebibliography}
\end{document}